\documentclass{aa}
\usepackage{graphicx}
\usepackage{txfonts}
 \begin{document}

   \title{Stability of latitudinal differential rotation in stars}

   \author{L.\,L.~Kitchatinov\inst{1,2} \and G.~R\"udiger\inst{1}
          }


   \institute{Astrophysikalisches Institut Potsdam, An der Sternwarte 16,
              D-14482, Potsdam, Germany \\
              \email{lkitchatinov@aip.de; gruediger@aip.de}
         \and
             Institute for Solar-Terrestrial Physics, PO Box
             291, Irkutsk 664033, Russia \\
             \email{kit@iszf.irk.ru}
             }

   \date{Received ; accepted }
  \abstract
 {}
 {The question is addressed whether stellar differentially rotating
radiative zones (like the solar tachocline) excite nonaxisymmetric
r-modes which can be observed. To this end the hydrodynamical
stability of latitudinal differential rotation is studied. The
amount of rotational shear required for the instability is estimated
in dependence of the character of radial stratification and the flow
patterns excited by the instability are found.}
 {The eigenvalue equations for the nonaxisymmetric disturbances
are formulated in 3D and then solved numerically. Radial
displacements and entropy disturbances are included. The equations
contain the 2D approximation of strictly horizontal displacements as
a special limit.}
 {The critical  magnitude of the latitudinal differential rotation
for onset of the instability is considerably reduced in the 3D
theory compared to the 2D approximation. The instability requires a
subadiabatic stratification. It does not exist in the bulk of
convection zone with almost adiabatic stratification but may switch
on near its base in the region of penetrative convection. Growth
rates and symmetry types of the  modes are computed in dependence on
the rotation law parameters. The S1 mode with its transequatorial
toroidal vortices is predicted as the dominating instability mode.
The  vortices show longitudinal drift rates  retrograde to the basic
rotation which  are close to that of the observed weak r-mode
signatures at the solar surface.}
  {}


   \titlerunning{Stability of stellar differential rotation}

   \authorrunning{L.\,L.~Kitchatinov \& G.~R\"udiger}

   \maketitle
\section{Introduction}
The differential stellar rotation may excite other types of motion
via instability. The possible transmission of rotational energy to
other types of motion may be relevant for various astrophysical
processes. Excitation of r-modes of global oscillations in
differentially rotating neutron stars is considered as a source of
detectable gravitational waves (Watts et al. \cite{Wea03}). Knaack
et al. (\cite{KSB05}) interpreted the large-scale structures in
magnetic fields of the Sun as signatures of r-modes which may in
turn result from an instability.

The stability problem of differential rotation is also relevant for
the  dynamics of the solar tachocline (Gilman \cite{G05}). The
tachocline is the  thin shell beneath the convection zone  where the
rotation pattern changes strongly. Beneath the tachocline the solar
rotation is practically uniform. Above the tachocline, the rotation
rate varies with latitude as observed at the solar surface. Inside
the tachocline, a transition from differential to uniform rotation
occurs with increasing depth. The question is whether this
tachocline is hydrodynamically stable or not. If it is, the idea
that the site of the solar dynamo is beneath the convection zone is
hard to accept.

The tachocline thickness is about 4\% of the solar radius.  The
tachocline is located mainly if not totally beneath the base of the
convection zone at $R_\mathrm{in} = 0.713R_\odot$
(Christensen-Dalsgaard et al. 1991; Basu \& Antia 1997) in the
uppermost radiation zone.

The stability/instability of the solar tachocline is also related
closely to the lithium problem. The lithium at the surfaces of cools
MS stars slowly decays with a characteristic time $\sim$1~Gyr. The
primordial lithium is destroyed at temperatures greater than 2.7 Mio
K which is already exceeded 42.000 km beneath the bottom of the
convection zone. Evidently, the tachocline should not be too
unstable otherwise the downward transport of the lithium may easily
be to strong. Nevertheless, the diffusion coefficient for the
lithium must exceed the molecular diffusion by one or two orders of
magnitude.

The hydrodynamical stability problem has  extensively been studied
in 2D approximation of purely toroidal disturbances. Symmetry types
and growth rates of the 2D unstable  modes are known (Watson
\cite{W81}; Dziembowski \& Kosovichev \cite{DK87}; Charbonneau et
al. \cite{CDG99}), and also the weakly nonlinear evolution  of the
instability  has been  described (Garaud \cite{G01}; Cally
\cite{C01}).  The 2D approximation neglects the radial displacements
which can be  expected as  small in stably stratified radiative
shells where the buoyancy frequency $N$ is much higher than
 the rotation rate, $N \gg \Omega$ (Watson
\cite{W81}). This  condition is not fulfilled in   stellar convection
zones.

The present paper overcomes the 2D approximation  by allowance for
radial displacements. Poloidal motions and entropy disturbances are
thus included. Our formulation contains the 2D approximation as a
special limit of large parameter $\hat{\lambda} = \ell N/(r\Omega )
\gg 1$; where $\ell$ is the radial scale of the disturbances and $r$
is the radius. We shall see that the most unstable modes have so
small radial scales that $\hat{\lambda}\lower.4ex\hbox{$\;\buildrel
<\over{\scriptstyle\sim}\;$}1$ and this condition is by far not
fulfilled. The minimum amount of differential rotation for onset of
the instability is considerably reduced compared to the 2D case.
More important is that the instability does not exist in the limit
of $N\rightarrow 0$, so that differential rotation is stable in
convection zones with their almost adiabatic stratification. The
instability may, however, switch on in the region of penetrative
convection near the base of the convection zone. Such a near-base
instability may be the reason for the difference in latitudinal
profiles of angular velocity between the top and the bottom of the
solar convection zone (Charbonneau et al. \cite{CDG99}). If this is
the case, transequatorial vortices (unstable S1 modes) should be
present near the base. The rates of (retrograde) drift of the
vortices are similar to that of the r-modes signatures inferred by
Knaack et al. (\cite{KSB05}) from solar magnetograms.
\section{The model}
The latitudinal dependence of the angular velocity $\Omega$ on the
sun can be approximated by an expression including $\cos^2\theta$
and $\cos^4\theta$ terms so that
 \begin{equation}
    \Omega = \Omega_0\left( 1 - a\left( (1-f)\cos^2\theta +
    f\cos^4\theta \right)\right) .
    \label{1}
 \end{equation}
Here  $\Omega_0$ is the equatorial angular velocity, $a$ is the
normalized equator-pole difference in the rotation rate, and  $f$ is
the fraction of the $\cos^4\theta$ term contribution to that
differential rotation. The rotation is assumed slow enough not to
deform the spherical symmetry of the star.

At  the solar surface it is $a\simeq 0.3$  and $f\simeq 0.5$ (Howard
et al. \cite{HAB83}). The latitudinal shear  varies only slightly
with depth in the bulk of the convection zone but it shows a
characteristic change near its base (Charbonneau et al.
\cite{CDG99}). The amplitude $a(1-f)$ of the $\cos^2\theta$ term
remains almost constant up to the base and starts decreasing in the
deeper tachocline only while the fraction $f$ of $\cos^4\theta$
contribution drops to practically zero near the base (cf. Fig.~10 of
Charbonneau et al. \cite{CDG99}).

The stratification is characterized by the buoyancy frequency $N$,
 \begin{equation}
   N^2 = \frac{g}{C_\mathrm{p}}\frac{\partial s}{\partial r} ,
   \label{2}
 \end{equation}
where $g$ is the gravity, $c_\mathrm{p}$ is the specific heat at
constant pressure, and $s$ is the specific entropy.

We address the linear stability problem with the small disturbances
depending on time as $\mathrm{exp}(-\mathrm{i}\omega t)$. A positive
imaginary part of the eigenvalue $\omega$ means an instability. The
radial scales of the disturbances are assumed small compared to the
stellar radius while the equations are  global in both the
horizontal dimensions. The dependencies on radius and longitude
$\phi$ are taken as Fourier modes $\mathrm{exp}(\mathrm{i}m\phi +
\mathrm{i}kr)$.

\subsection{Equations}\label{eqs}
The linear equations for  small perturbations in differentially
rotating fluids with toroidal magnetic fields are already given by
Kitchatinov \& R\"udiger (\cite{KR08}). Here, the nonmagnetic
version of the equations is considered in detail. The equations are
formulated for normalized parameters (the rules of conversion to
physical variables are given below). The equation for the potential
$W$ of the toroidal flow reads
  \begin{eqnarray}
    (\hat{\omega} &-& m \hat{\Omega} ) \left(\hat{L}W\right)\ =
    -\mathrm{i}\frac{\epsilon_\nu}{\hat{\lambda}^2} \left(\hat{L}W\right) -
    m \frac{\partial^2\left((1-\mu^2)\hat{\Omega}\right)}{\partial\mu^2}\ W
    \nonumber \\
    &+& \frac{\partial\left((1-\mu^2)\hat{\Omega}\right)}{\partial\mu}
    \left(\hat{L}V\right) +
    \frac{\partial^2\left((1-\mu^2)\hat{\Omega}\right)}{\partial\mu^2}\
    (1-\mu^2)\frac{\partial V}{\partial\mu} ,
    \label{3}
 \end{eqnarray}
where $\hat{\omega} = \omega/\Omega_0$ is the normalized eigenvalue,
$\hat{\Omega} = \Omega/\Omega_0$ is the normalized rotation rate, $\mu =
\cos\theta$, $V$ is the poloidal flow potential,
\begin{equation}
    \hat{L} = \frac{\partial}{\partial\mu}\left(1-\mu^2\right)
    \frac{\partial}{\partial\mu} - \frac{m^2}{1-\mu^2}\ \
    \label{4}
 \end{equation}
is the angular part of the Laplacian operator, and
\begin{equation}
    \hat{\lambda} = \frac{N}{\Omega_0 kr}\ \
    \label{5}
 \end{equation}
is the key parameter for the influence of the  stratification. The
diffusion terms are characterized by  the parameters
\begin{equation}
    \epsilon_\nu = \frac{\nu N^2}{\Omega_0^3 r^2},\ \ \ \ \ \ \ \ \ \ \ \ \ \ \ \ \ \ \ \ \
    \epsilon_\chi = \frac{\chi N^2}{\Omega_0^3 r^2},
    \label{6}
 \end{equation}
where $\nu$ and $\chi$ are the microscopic viscosity and the thermal conductivity.

Apart from the Eq.~(\ref{3}) for the toroidal flow, the complete system
of three equations includes the equation for poloidal flow,
\begin{eqnarray}
    (\hat{\omega} &-& m \hat{\Omega}) \left(\hat{L}V\right)\ =
    -\mathrm{i}\frac{\epsilon_\nu}{\hat{\lambda}^2} \left(\hat{L}V\right) -
    \hat{\lambda}^2 \left(\hat{L}S\right) +
    \nonumber \\
    &+& 2 m\left(\frac{\partial (\mu\hat{\Omega})}{\partial\mu}\ V + (1 -
    \mu^2)\frac{\partial\hat{\Omega}}{\partial\mu}\frac{\partial
    V}{\partial\mu}\right) -
    \nonumber \\
    &-& 2\mu\hat{\Omega}\left(\hat{L}W\right) - 2
    (1-\mu^2)\frac{\partial(\mu\hat{\Omega})}{\partial\mu}
    \frac{\partial W}{\partial\mu} - 2 m^2
    \frac{\partial\hat{\Omega}}{\partial\mu}\ W ,
    \label{7}
 \end{eqnarray}
and the equation for the normalized entropy $S$,
 \begin{equation}
    (\hat{\omega} - m \hat{\Omega})\ S = -\mathrm{i}
    \frac{\epsilon_\chi}{\hat{\lambda}^2} S  +
    \hat{L}V .
    \label{8}
 \end{equation}

The equations (\ref{3}), (\ref{7}) and (\ref{8}) form an eigenvalue
problem which we  solved numerically. The diffusion parameters
(\ref{6}) for the upper radiative core of the Sun are $\epsilon_\chi
= 10^{-4}$ and $\epsilon_\nu = 2\times 10^{-10}$ which are used in
the calculations.

The disturbances in physical units follow  from their normalized
values by
 \begin{equation}
    s = -\frac{\mathrm{i} C_\mathrm{p} N^2}{gkr}\ S,\ \ \ P_u =
    \left(\Omega_0 r^2/ k\right)\ V,\ \ \ T_u = \left(\Omega_0
    r^2\right)\ W.
    \label{9}
 \end{equation}
The velocity field can be restored from the potentials of poloidal
($P_u$) and toroidal ($T_u$) flows,
\begin{eqnarray}
    \vec{u} &=& \frac{\vec{e}_r}{r^2} \hat{L}P_u -
    \frac{\vec{e}_\theta}{r}
    \left(\frac{1}{\sin\theta}\frac{\partial T_u}{\partial\phi} +
    \frac{\partial^2 P_u}{\partial r\partial\theta}\right)
    \nonumber \\
    && +\ \frac{\vec{e}_\phi}{r}\left(\frac{\partial T_u}{\partial\theta}
    - \frac{1}{\sin\theta}\frac{\partial^2 P_u}{\partial
    r\partial\phi}\right)
    \label{10}
 \end{eqnarray}
(Chandrasekhar \cite{C61}), where $\vec{e}_r$, $\vec{e}_\theta$, and
$\vec{e}_\phi$ are unit vectors in the radial, meridional and
longitudinal directions.

Without rotation ($\Omega\rightarrow 0$) and for small diffusion
the equations (\ref{3}), (\ref{7}) and (\ref{8}) reproduce the
spectrum
 \begin{equation}
    \omega^2 = \frac{l(l+1) N^2}{r^2k^2},\ \ \ l = 1,2,...
    \label{11}
 \end{equation}
of $g$-modes. More related to the stability problem is the  limit of very large
$\hat\lambda$-parameter (\ref{5}) which leads to the following
2D approximation.

\subsection{2D approximation}\label{2D}
The ratio of $N^2/\Omega^2$ in stars can be so large ($\sim 10^5$ in
the upper radiative core of the Sun) that $\hat{\lambda}^2$
(\ref{5}) can also be large in spite of short-wave approximation in
radius, $kr \gg 1$.  In the limit of large $\hat{\lambda}^2$ the
above equation system reduces to its 2D approximation. In leading
order of this parameter Eq. (\ref{7}) gives $S = 0$. Then it follows
from (\ref{8}) that $V = 0$ and Eq.  (\ref{3}) reduces to the
standard equation of 2D theory of Watson (\cite{W81}),
 \begin{equation}
    \left(\hat{\omega} - m\hat{\Omega}\right)\left(\hat{L}W\right) =
    - m \frac{\partial^2\left((1-\mu^2)\hat{\Omega}\right)}
    {\partial\mu^2}\ W ,
    \label{12}
 \end{equation}
describing toroidal flows on spherical surfaces.

The 2D approximation is justified for stable oscillations with not
too short radial scales so that $\hat{\lambda}$ remains large. Its
validity for stability problem is less certain because the radial
scales of most rapidly growing modes are not know in advance and the
value of $kr$ for those modes is normally so large that
$\hat{\lambda}\lower.4ex\hbox{$\;\buildrel<\over{\scriptstyle\sim}\;$}
1$ (Kitchatinov \& R\"udiger \cite{KR08}).

For  rigid rotation Eq. (\ref{12}) provides the
 eigenvalue spectrum
\begin{equation}
    \omega = m\Omega \left( 1 - \frac{2}{l(l+1)}\right),\ \ \ l =
    1,2,...
    \label{13}
\end{equation}
of the r-modes (Papaloizou \& Pringle \cite{PP78}). Instabilities
can emerge with nonuniform rotation. The {\em necessary} condition
for instability is that the second derivative,
$\mathrm{d}^2((1-\mu^2)\hat{\Omega})/\mathrm{d}\mu^2$, changes its
sign (Watson \cite{W81}). For the angular velocity profile
\begin{equation}
    \hat{\Omega} = 1 - a\mu^{2n}
    \label{14}
 \end{equation}
the condition demands
 \begin{equation}
    a > \frac{1}{4n+1} .
    \label{15}
 \end{equation}
It should be $a > 0.2$ for $n=1$. For $n=2$, i.e. the
$\cos^4\theta$-profile (\ref{14}), the amplitude $a$ of differential
rotation for the onset of instability may reduce considerably, $a >
1/9$. This is why the $\cos^4\theta$-term is kept in the
differential rotation profile (\ref{1}).

The profile (\ref{1}) for the Sun is, of course, an approximation.
Higher order terms in $\cos^2\theta$ may also be present. Reduction
of the instability threshold due to the higher order terms is,
however, less significant and they are relevant only to the
near-polar regions. Here, the results are presented for the rotation
law (\ref{1}).

\subsection{Symmetry types}
The eigenmodes provided by both the 2D approximation (\ref{12}) and
the full 3D equation system of section~\ref{eqs} possess definite
equatorial symmetries. We  use the notation Sm for the modes with
symmetric relative to the equator potential $W$ of toroidal flow and
the notation Am for antisymmetric $W$ (m is the azimuthal wave
number). The symmetry convention is same as used before (Charbonneau
et al. \cite{CDG99};  Kitchatinov \& R\"udiger \cite{KR08}). Note
that the eigenmodes combine $W$ of definite symmetry with $S$ and
$V$ of opposite symmetry, i.e., $S$ and $V$ are symmetric for Am
modes and antisymmetric for Sm. The velocity field for Am modes has
antisymmetric $u_\theta$ and symmetric $u_r$ and $u_\phi$ about the
equatorial plane, and the other way round for Sm modes.

Watson (\cite{W81}) proved that only nonaxisymmetric modes with
$m=1$ and $m=2$ can be unstable in the 2D approximation. Our 3D
computations yield the same conclusion.

\section{Results}
\begin{figure}
   \centering
   \includegraphics[width=7.0cm]{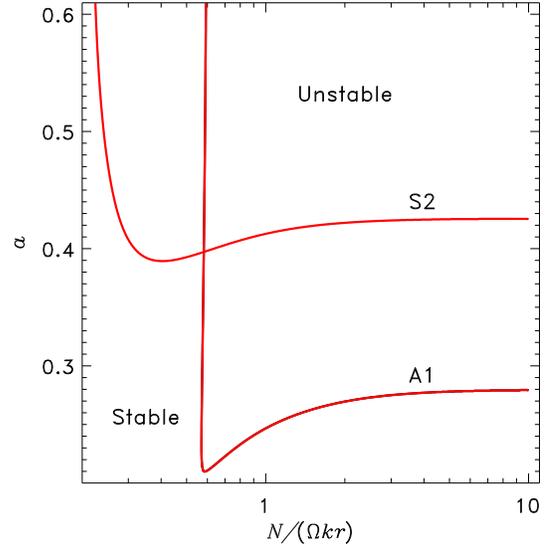}
   \caption{Neutral stability lines for  $f=0$ in rotation
    law (\ref{1}). The instability region is above the lines. Only A1 and S2 modes are  unstable.
    The lines approach the marginal $a$-values of Watson theory for large $\hat{\lambda}$.
    Note that the most unstable  modes have $\hat{\lambda} < 1$.
              }
   \label{f1}
\end{figure}

Figure~\ref{f1} shows the critical shear amplitudes $a$ as functions
of $\hat{\lambda}$ for $f=0$. This is the case considered by Watson.
As it must be, his results are reproduced in the limit of large
$\hat{\lambda}$. Note, however,  that the most easily excited modes
have $\hat{\lambda} < 1$.

Even for large $N/\Omega$, the small radial displacements are
significant for the instability. The reason is that the most
unstable modes have short radial scales. It can be seen from
(\ref{10}) that zero radial velocity  would exclude  the whole class
of poloidal disturbances. The ratio of horizontal ($u_\mathrm{h}$)
to radial velocities in a cell of poloidal flow of different radial
($\ell$) and horizontal ($H$) scales can be estimated as
$u_\mathrm{h}/u_r \sim H/\ell$. Horizontal velocity of poloidal flow
can thus remain important  in spite of small $u_r$ if the radial
scale is much shorter than the horizontal one. The poloidal
(interchange-type) disturbances are so significant  for the
instability that the critical latitudinal shear  for onset of the
instability reduces from $a=0.28$ (value of Watson) to $a=0.21$.
However, this smaller value still ensures the tachocline to be
stable.

\begin{figure}
   \centering
   \includegraphics[width=7.0cm]{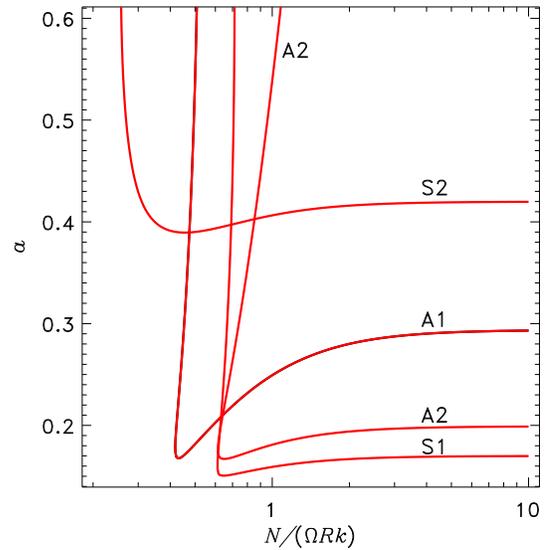}
   \caption{The same as in Fig. \ref{f1} but for $f=0.5$.
    The critical shear for onset of the instability is reduced,
    the newly appeared unstable modes S1 and A2 are excited most easily.
              }
   \label{f2}
\end{figure}

We find that the instability disappears when  the
stratification approaches adiabaticity ($N\rightarrow 0$). The
instability of differential rotation does thus {\em not} exist in convection
zones.

\begin{figure}
   \centering
   \resizebox{\hsize}{!}{\includegraphics{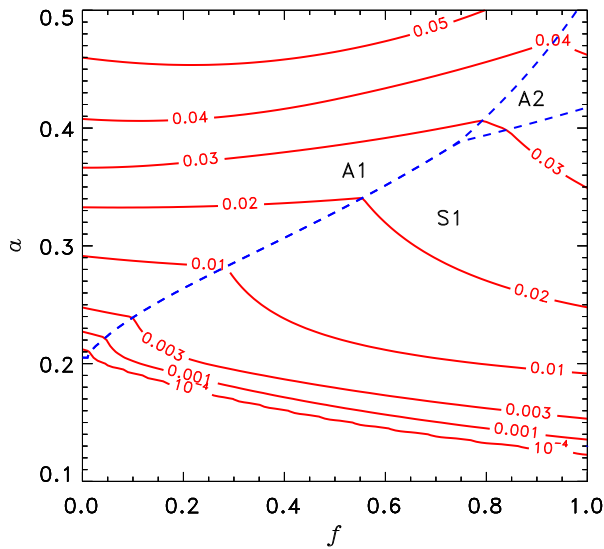}}
   \caption{Isolines for the  normalized growth rates $\sigma = \Im (\omega )/\Omega_0$ of most
    rapidly growing modes. Dominating symmetry types are indicated.
    Dashed lines show the watershed between different symmetry types.
              }
   \label{f3}
\end{figure}

Fig.~\ref{f2} reveals  a striking destabilizing effect of finite $f$
in the rotation law (\ref{1}). The threshold value of $a$ for
marginal stability is further reduced compared to the case of $f=0$.
New unstable modes appear and the S1 mode is now preferred.

We find the  instability as rather sensitive to the details of the
rotation law. The growth rates of the unstable modes in dependence
on $a$ and $f$ are shown in Fig.~\ref{f3}. The length scale
$\hat{\lambda}$ was varied to find the maximum growth rates shown in
the plot. The dashed line separates the regions of different
symmetry types. Surprisingly, even the symmetries of the most
rapidly growing modes depend on the shape of the rotation law. The
shape of the rotation law is the result of the interaction of the
turbulence and the basic  rotation in the solar/stellar convection
zone.
\subsection{Angular momentum transport}
\begin{figure}
   \centering
   \includegraphics[width=8.0cm]{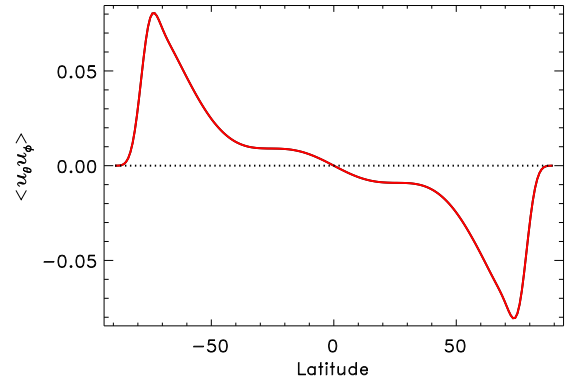}
   \caption{Meridional flux of angular momentum for slightly supercritical
    ($\sigma = 0.001$) S1-mode for  $f=0.2$ with normalized velocities.
              }
   \label{f4}
\end{figure}

The depth dependence of the solar rotation law known from
helioseismology can be interpreted in light of the presented
results. The rotation law in the bulk of the almost adiabatically
stratified convection zone is stable. In the region of penetrative
convection near the base the stratification changes to
subadiabaticity and the instability can exist. If it exists, then it
reacts back on the differential rotation to change it towards a
stable profile with $f=0$. Our linear computations cannot describe
this  nonlinear  process but they can probe for the sense of angular
momentum transport. Figure~\ref{f4} shows that the instability
indeed tends to reduce the differential rotation. The plot shows the
angular momentum flux  $\langle u_\theta u_\phi \rangle$   after
longitude-averaging as a function of the latitude. The correlation
is negative (positive) for the northern (southern) hemisphere. The
angular momentum is thus transported from the equator to the poles.
The plot was constructed for slightly supercritical S1 mode which
should be active if the differential rotation is reduced to the
marginally stable value (Fig.~\ref{f3}).
\subsection{The flow pattern}
\begin{figure}
   \centering
   \includegraphics[width=8.0cm]{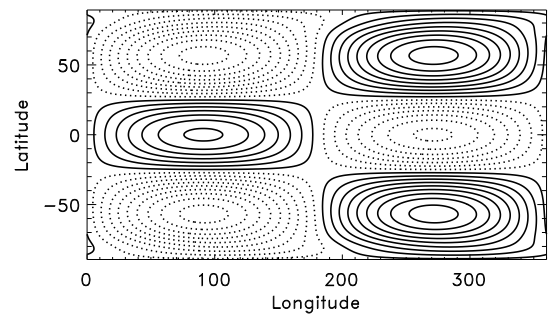}
   \caption{Streamlines of toroidal flow for the same mode as in Fig.~\ref{f4}. Full
    and dotted lines show opposite senses of circulation.
              }
   \label{f5}
\end{figure}

\begin{figure}
   \centering
   \includegraphics[width=7.0cm]{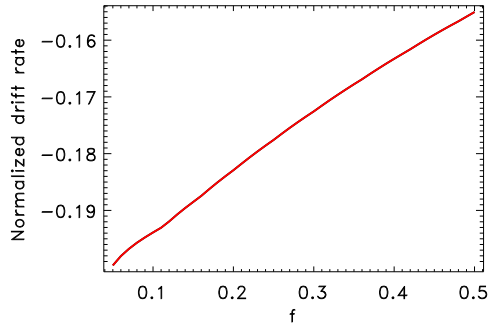}
   \caption{Normalized drift rates $\Re (\omega )/\Omega_0 -1$ in
    corotating frame for slightly supercritical ($\sigma = 0.001$) S1-modes
    as function of $f$-parameter of rotation law (\ref{1}).
              }
   \label{f6}
\end{figure}

The streamlines of the toroidal flow for the same S1 mode are shown in
Fig.~\ref{f5}. Close to the equator the flow represents
transequatorial vortices. The flow pattern drifts in longitude
against the direction of rotation in the corotating frame with   rates
 shown in Fig.~\ref{f6}. If the deep-seated vortices were
observable (e.g., due to disturbance of the large-scale magnetic
field), the observer  would  see
the frequencies
 \begin{equation}
     \nu = m\left(\nu_\odot - \nu_\mathrm{E}\right) + \nu_\odot\ \hat{\nu} ,
    \label{16}
 \end{equation}
where $\nu_\odot = 450$\,nHz is the equatorial rotation
frequency of the base of the convection zone, $\nu_\mathrm{E} =
31.7$\,nHz is the orbital rotation frequency of the Earth, and
$\hat{\nu}$ is normalized drift rate of Fig.~\ref{f6}. The equation
gives the range of frequencies from about 120 to 160 nHz for the
range of drift rates of Fig.~\ref{f6}. The frequencies correspond to
periods from about 120 to 160 days for the method of analysis of
synoptic maps of solar magnetic fields by Knaack et al.
(\cite{KSB05}). Two of the periods which Knaack et al. interprete
as signatures of r-modes fall into this range.
\section{Discussion}
Rotation laws in stably stratified stellar interiors with strong
enough latitudinal gradient are hydrodynamically unstable against
nonaxisymmetric disturbances or -- what is the same -- the r-modes
are excited for large enough differential rotation.  As their drift
rate is retrograde with an amplitude of about 10\% of the rotation
rate they should be observable.

They are only excited, however, in subadiabatically stratified
radiative zones and convective overshoot regions. Hence, their
existence directly indicates the extended regions of latitudinal
shear in the deep stellar interior beneath the proper convection
zone. If this region -- like the solar tachocline  -- is thin the
intensity of the r-mode oscillation should be much smaller than in
stars without tachoclines. New calculations for rotation laws with
radial profiles of the angular velocity  are necessary to develop
this new r-mode seismology. In the present paper the radial shear of
rotation is not included that can be justified only if the angular
velocity varies in radius on a scale larger than the radial
wavelength $\Omega r/N$ of unstable excitations.

The critical latitudinal shear for the excitation of the modes is
not very small. The simplest theory without radial perturbations and
with a simplified  parabolic rotation law yields a  critical
latitudinal shear of 28\%. We have shown with an improved
mathematics that the real value is less. It is reduced to 21\% for
the same rotation law but with a 3D theory. The critical shear value
is further reduced if the rotation law contains a higher order term
of $\cos^4\theta$. Nevertheless, the critical shear rate remains
larger than (say) 10\%. A rotation law with a slightly smaller
latitudinal shear (driven by the turbulence in the convection zone)
could stably exist in the stellar interior without any  decay. We
know, however, that the solar core rotates almost rigidly.  This is
only understandable with  the idea that the slender  solar
tachocline is due to an extra effect, e.g. by the Maxwell stress of
large-scale magnetic fields.  They may be of fossil origin as their
amplitudes need not exceed (say) 1 Gauss (cf. R\"udiger \&
Kitchatinov \cite{RK07}).

If this magnetic concept for the solar tachocline is true and if the
tachocline is indeed hydrodynamically stable then also the very slow
decay of the observed lithium abundance becomes understandable. The
slight increase of the diffusion  coefficient by one or two orders
of magnitude can easily be explained with  slow horizontal motions
of order cm/s (see R\"udiger \& Pipin \cite{RP01}) or by radial
plumes overshooting from the convection zone  (Bl\"ocker et al.
\cite{Bea98}).
\begin{acknowledgements}
This work was supported by the Deutsche
For\-schungs\-ge\-mein\-schaft and by the Russian Foundation for
Basic Research (project 09-02-91338).
\end{acknowledgements}

\end{document}